\documentclass[11pt, a4paper, onecolumn]{deepmind}

\pdftrailerid{redacted}

\makeatletter
\renewcommand\bibentry[1]{\nocite{#1}{\frenchspacing\@nameuse{BR@r@#1\@extra@b@citeb}}}
\makeatother

\usepackage{hyperref}

\usepackage{multirow}
\usepackage{caption}
\usepackage{subcaption}
\usepackage{flushend}
\usepackage{kantlipsum, lipsum}
\usepackage{dsfont}
\usepackage{dm-colors}
\usepackage{nicefrac}

\usepackage{todonotes}
\usepackage{float}
\usepackage{enumitem}
\usepackage{cleveref}

\usepackage{url}
\usepackage{breakurl}

\usepackage{tocloft}
\usepackage{etoc}
\usepackage{setspace}
\usepackage{multicol}
\AtBeginDocument{\addtocontents{toc}{\protect\thispagestyle{firststyle}}} 

\usepackage[authoryear, sort&compress, round]{natbib} 

\graphicspath{{figures/}}

\newcommand{\Copy}{\texttt{Copy}}
\newcommand{\NoCopy}{\texttt{NoCopy}}
\newcommand{\Drop}{\texttt{Drop}}
\newcommand{\mallocGame}{\emph{MMapGame}}
\newcommand{\mallocMuZero}{\emph{MMap-MuZero}}
\newcommand{\prodmallocMuZero}{\emph{MMap-MuZero-prod}}
\newcommand{\dropbackup}{$\Drop$-\emph{backup}}

\newtheorem*{theorem*}{Theorem}

\let\originalparagraph\paragraph
\renewcommand{\paragraph}[2][.]{\originalparagraph{#2#1}}

\title{Optimizing Memory Mapping Using Deep Reinforcement Learning}

\correspondingauthor{pengming@poolside.ai, sazanovich@google.com}

\reportnumber{001} 

\author[*, 3]{Pengming Wang}
\author[*, 1]{Mikita Sazanovich}
\author[2]{Berkin Ilbeyi}
\author[1]{Phitchaya Mangpo Phothilimthana}
\author[2]{Manish Purohit}
\author[2]{Han Yang Tay}
\author[1]{Ngân Vũ}
\author[1]{Miaosen Wang}
\author[1]{Cosmin Paduraru}
\author[1]{Edouard Leurent}
\author[3]{Anton Zhernov}
\author[1]{Po-Sen Huang}
\author[1]{Julian Schrittwieser}
\author[1]{Thomas Hubert}
\author[3]{Robert Tung}
\author[1]{Paula Kurylowicz}
\author[1]{Kieran Milan}
\author[1]{Oriol Vinyals}
\author[1]{Daniel J. Mankowitz}

\affil[*]{Corresponding author}
\affil[1]{Google DeepMind}
\affil[2]{Google}
\affil[3]{Contributions while at Google DeepMind}

\begin{abstract}
Resource scheduling and allocation is a critical component of many high impact systems ranging from congestion control to cloud computing. Finding more optimal solutions to these problems often has significant impact on resource and time savings, reducing device wear-and-tear, and even potentially improving carbon emissions. In this paper, we focus on a specific instance of a scheduling problem, namely the memory mapping problem that occurs during compilation of machine learning programs: That is, mapping tensors to different memory layers to optimize execution time.

We introduce an approach for solving the memory mapping problem using Reinforcement Learning. RL is a solution paradigm well-suited for sequential decision making problems that are amenable to planning, and combinatorial search spaces with high-dimensional data inputs. We formulate the problem as a single-player game, which we call the \mallocGame, such that high-reward trajectories of the game correspond to efficient memory mappings on the target hardware. We also introduce a Reinforcement Learning agent, \mallocMuZero, and show that it is capable of playing this game to discover new and improved memory mapping solutions that lead to faster execution times on real ML workloads on ML accelerators. We compare the performance of \mallocMuZero\ to the default solver used by the Accelerated Linear Algebra (XLA) compiler on a benchmark of realistic ML workloads. In addition, we show that \mallocMuZero\ is capable of improving the execution time of the recently published AlphaTensor matrix multiplication model.

\end{abstract}

\begin{document}

\maketitle

\onecolumn
\titleformat{\part}
 {}
 {}
 {0pt}
 {\color{black}}
\vspace{-1.6cm} 
\part{}
{
\setlength{\columnseprule}{0.5pt}
\begin{multicols}{2}
\etocsettocstyle{}{}
\etocsetnexttocdepth{subsection}
\localtableofcontents
\end{multicols}
}

\onecolumn

\newpage
\section{Introduction}
\label{sec:introduction}

Compute resource efficiency is critical in today's large-scale systems, and plays an increasingly greater role as demand for compute increases. In particular in the domain of machine learning, the demand for increased compute is accelerating at a fast pace, as workloads grow larger, and applications proliferate. Improving resource efficiency for ML workloads hence presents an important opportunity. One promising avenue towards this goal is to improve ML compilers to optimize ML programs to utilize hardware more efficiently, as proposed in e.g. \citep{autotuner,telamalloc, chen2018learning, chen2018tvm, li2020adatune, jia2019taso, steiner2021value}. 

\begin{figure*}[ht]
    \centering
    \includegraphics[width=0.5\linewidth]{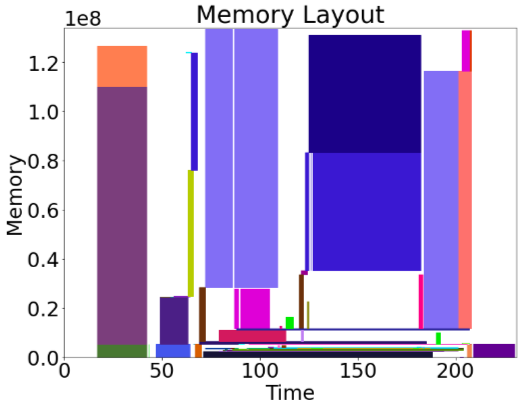}
    \caption{An example memory mapping for the \emph{AlexNet} model on a TPUv4i. Each rectangle represents the memory reserved in CMEM for a particular tensor output or operand: The x-axis represents the logical time of the program, while the y-axis represents memory space. Outputs/operands with the same colour represent the same tensor in the program.}
    \label{fig:mem_layout}
\end{figure*}

In the same vein, we focus in this paper on the \emph{memory mapping} problem. Modern hardware architectures have multiple layers of memory hierarchy, differing in their sizes and speeds; typically ranging from large, but slow memory layers (e.g. HBM on TPUv4), to increasingly smaller, but faster layers (e.g. CMEM on TPUv4). We call the problem of determining when to use which memory layer, and managing data transfer between layers, the \emph{memory mapping} problem.
More specifically, a solution to the memory mapping problem defines exactly which buffers are allocated at what offsets in the fast memory, as well as the time interval each buffer is allocated in memory. This can be visualized as a 2-dimensional image, e.g. as in Figure~\ref{fig:mem_layout}, depicting for each buffer assigned to fast memory its memory offset and time interval. A good memory mapping means that the faster memory layers are utilized effectively, which can significantly reduce the overall execution time of the program.

Finding optimal, or even just good solutions is an extremely challenging problem, as one needs to balance the resource trade-offs between fast memory space, execution time, and inter-memory bandwidth used for prefetching. This can be seen as an NP-hard scheduling problem. In practice, the memory mapping problem is typically solved by compilers such as XLA~\citep{xla} through a series of expert-designed rule-based heuristics. While these approaches often perform well on average, they also frequently yield suboptimal results, as a fixed set of rules can not cover all complex cases. Instead, we introduce an approach that uses reinforcement learning to solve this problem, using the power of search and learning to find more optimal mappings for ML programs.

To frame the memory mapping problem as an RL problem, we introduce a single-player game which we refer to as the \mallocGame. In this game, a player receives as input a program as a sequence of instructions where each instruction has a set of tensor outputs or operands. The player determines whether to place each tensor output/operand into a limited size, but fast memory (e.g. CMEM) or into a larger, but slow memory (e.g. HBM) with the goal of optimizing the total execution time of the program. Each tensor output/operand has a pre-defined memory size and execution time. For each output/operand, the player needs to decide whether or not to allocate it in fast memory, and whether or not to schedule data transfer between fast and slow memories. The decisions are subject to hardware constraints, such as the size of fast memory, or the data transfer bandwidth between memories. Through these decisions for each buffer, the game incrementally builds a solution for the memory mapping problem.

This is a very challenging problem for a number of reasons. Firstly, as fast memory is limited, it is typically not possible to serve all instructions from there. In addition, the copy bandwidth between fast and slow memories is limited, and moving buffers between memories can add additional execution overhead. As such, the player needs to balance the trade-off between available memory space, copy bandwidth and execution time efficiently. In addition, a program can have up to $10^4$ instructions which can make a single game trajectory very long. This results in an extremely large, combinatorial search space of over $10^{4000}$ possible game trajectories, exceeding other challenging games such as Chess ($10^{120}$ trajectories)~\citep{silver2018general} or Go ($10^{700}$ trajectories)~\citep{silver2016mastering}.

In this game, early decisions have long-lasting consequences. For instance, blocking memory space that could be used more efficiently later on or taking up too much copy bandwidth to copy an important buffer into CMEM, could lead to sub-optimal performance results. As a result, planning is critical in this problem domain, so we introduce our Reinforcement Learning agent \mallocMuZero, an extension of the well-known MuZero agent \citep{schrittwieser2020mastering} that plays the \mallocGame. This agent utilizes a novel representation network to understand the structure of the memory allocation problem at hand and also incorporates a Drop-\textit{backup} mechanism that enables it to handle infeasible states.

We apply this approach to optimize memory mapping for realistic ML workloads running on the TPUv4i ML accelerator, which features CMEM as a scratchpad fast memory, and HBM as the slow main memory. We integrate our approach with the XLA (Accelerated Linear Algebra) library~\citep{xla} compiler and evaluate the end-to-end latency of the compiled programs. We compare the resulting execution times with the XLA compiler using default settings.

\paragraph{Contributions}
In this paper, we formulate the memory mapping problem as a single-player game, which we refer to as \mallocGame, and introduce a Reinforcement Learning agent \mallocMuZero\ to play this game. \mallocMuZero\ extends MuZero \citep{schrittwieser2020mastering} with a domain specific representation network as well as a drop-\textit{backup} mechanism that prevents infeasible states from being encountered. This algorithm is trained and evaluated on realistic ML workloads including 52 models from the XLA benchmark, as well as 8 high-impact workloads from Alphabet's fleet. Our agent is able to improve upon 33 out of the 60 programs, with an average speedup over the entire benchmark of $0.59\%$ as a stand-alone agent, and a maximum speedup of $87\%$. \mallocMuZero\ is also able to achieve a memory mapping speedup of $5.78\%$ on a version of the recently published AlphaTensor model~\citep{alphatensor}. We also provide a set of investigative studies that provide further insight into the performance of our agent. Finally, we also introduce \prodmallocMuZero\, a hybrid agent that combines the policy of \mallocMuZero\ with the current production heuristic policy, reflecting how it would be incorporated into a production setup. This combined agent is able to achieve an average execution time improvement of $4.05\%$. This is a significant achievement, as even single percentage improvements represent large savings at scale.

\section{Related Work}
\label{sec:related_work}

\paragraph{ML compiler optimization} There are many optimization problems that are solved during the compilation of ML workloads, many of which are areas of active research, such as device placement (deciding on which device to execute which operations)~\citep{paliwal2019regal, placement}, scheduling (when to execute which operation)~\citep{zhou2020transferable}, fusion (deciding which operations to merge)~\citep{zhou2020transferable}, and memory allocation~\citep{telamalloc}. A general framework for autotuning compiler passes was also proposed in~\citep{autotuner}.

\paragraph{RL for scheduling} There is much prior work in using RL for scheduling tasks. Many of the solutions are based on using Reinforce \citep{reinforce}, Q-learning \citep{sutton2018reinforcement}, DQN \citep{dqn}, A3C \citep{a3c} or variants thereof to better schedule resources \citep{rss1,rss2,rss3,rss4,rss5,rss6,zhang1995reinforcement}. Search-based RL algorithms such as AlphaZero have also been previously used for production optimization \citep{azsheetmetal}. There is also vast amounts of literature on alternative optimization techniques for resource scheduling such as particle swarm optimization \citep{particleswarm, particleswarm2, particleswarm3} and supervised learning \citep{supervised}, as well as techniques that tackle different aspects of schedulers including fairness \citep{fairness,fairness2}. 

\section{Background}
\label{sec:background}

\paragraph{Markov Decision Process} \label{par:bg:mdp}
A Markov Decision Process (MDP) is defined as the $5$-tuple $(\mathcal{S}, \mathcal{A}, P, R, \gamma)$, where $\mathcal{S}$ is the \emph{state space}, describing the set of observable states to the agent; $\mathcal{A}$ is the \emph{action space}, describing the set of possible actions; $P:\mathcal{S}\times \mathcal{A} \rightarrow [0,1]^{\mathcal{S}}$ is the \emph{state transition function} describing the probability of transitioning to state $s_{t+1}$ given a state $s_t$ and an action $a_t$; $R:\mathcal{S}\times \mathcal{A} \rightarrow \mathbb{R}$ is a bounded \emph{reward function}; and $\gamma\in [0, 1]$ is the \emph{discount factor}. The solution to an MDP is a policy $\pi:\mathcal{S}\rightarrow \Delta_{\mathcal{A}}$ which maps a given state to a distribution over actions. The goal is then to find a policy $\pi$ with maximal value $V^\pi$ at an initial state $s\in\mathcal{S}$, defined as the expected discounted cumulative reward $V^\pi(s) = \mathbb{E}\biggl[\sum_{t=0}^\infty \gamma^t R(s_t, a_t) \mid s_0 = s\biggr]$.

\paragraph{MuZero} \label{par:bg:muzero}
MuZero~\citep{schrittwieser2020mastering} is a model-based RL algorithm that leverages Monte-Carlo tree search (MCTS) as a policy improvement operator.
In contrast to its predecessor AlphaZero~\citep{silver2018general} which uses the true dynamics and rewards when planning, MuZero plans in a latent space by making use of three trainable models: (i) a \emph{representation network} $f^{\text{rep}}$ that outputs a latent representation $h_t$ of the state $s_t$; (ii) a \emph{dynamics network} $f^{\text{dyn}}$ that predicts the next latent state $h_t^{k+1}$ and reward $\hat{r}_{t}^{k+1}$ resulting from a transition. Note that the subscript $t$ denotes timesteps in the real environment and the superscript $k$ represents timesteps in the model; (iii) a \emph{prediction network} $f^{\text{pred}}$ that predicts the expected return (the value) $\hat{v}_t$ and a policy (\emph{i.e.} distribution over the action space) $\hat{\pi}_t$ from a given latent state.
\begin{align}
    h_{t} =&~ f^{\text{rep}}(s_t) \\
    h_{t}^{k+1},\,\hat{r}_{t}^{k+1} =&~ f^{\text{dyn}}(h_t^k, a_t^k) \\
    \hat{v}_t,\,\hat{\pi}_t =&~ f^{\text{pred}}(h_t)
\end{align}

Upon reaching a new state, MuZero proceeds by first encoding the state into a latent representation with the representation network.
Then, the dynamics network $f^{\text{dyn}}(h_t^k, a_t^k)$ and prediction network $f^{\text{pred}}(h_t)$ are used to simulate several trajectories that fill out a search tree, by sampling state transitions. At each node, the actions are selected with an optimistic strategy called \textit{Predictor Upper Confidence Tree} (PUCT) bound \citep{silver2016mastering}, meant to balance \textit{exploration} (trying new actions), and \textit{exploitation} (exploring further the subtree of the current estimate of the best action). This strategy starts out by following the predicted policy $\hat{\pi}_t$ closely, and gradually shifts towards maximising the predicted value function. Ultimately, an action is recommended by sampling from the root node with probability proportional to its visit count during MCTS. The predicted policy is then trained to match the visit counts of the MCTS policy, in an attempt to distill the search procedure into a policy such that subsequent iterations of MCTS will disregard nodes that are not promising.

\paragraph{Reanalyse} \label{par:bg:reanalyse}
To increase sample efficiency in MuZero, we can take advantage of external demonstrations. The idea of Reanalyse \citep{schrittwieser2021online} is to update the agent's model and prediction parameters based on data it has already experienced or a demonstration to yield improved training targets. The MCTS procedure generates fresh policy and value targets for each state of the demonstration. Reanalyse can be repeatedly applied to old trajectories to generate fresher and better targets as the training continues.

\paragraph{XLA} XLA~\citep{xla} is a domain-specific compiler designed to accelerate linear algebra and machine learning workloads on different hardware targets, including CPU, GPU, and TPU. It powers popular machine learning frameworks such as TensorFlow~\citep{tensorflow} and JAX~\citep{jax}. During compilation, XLA performs a series of analysis and optimisation processes which significantly impact the performance and resource efficiency of the compiled program. In this work, we address the \emph{memory mapping} component, which is the task of utilising different memory hierarchies efficiently. We specifically focus on the problem of managing the fast CMEM memory layer on TPU4i hardware, though the techniques described are general and can be applied to other architectures as well.

\section{Deep Reinforcement Learning for Memory Mapping}
\label{sec:drl}

\subsection{Memory Mapping Problem}
We first define the memory mapping problem in more formal terms. The input to the problem is a \emph{program} $\mathcal{P} = (\mathcal{I}_1, ..., \mathcal{I}_T)$ given as a sequence of $T$ \emph{instructions} $\mathcal{I}_i$. We refer to the indices in the instruction sequence as the \emph{logical time} of the program. Each instruction has a set of inputs and outputs, which we collectively call the \emph{buffers} used by the instruction. Each buffer has a set of properties, such as its size, the logical time of its instruction (its position in the instruction sequence of the program), or the expected speedup when reading (or writing) the buffer from fast memory (in our case CMEM) instead of HBM. The full list of buffer features can be found in Table~\ref{tab:features}. Furthermore, we are also supplied with the total size of CMEM available $\mathrm{max\_size}$ and assume that the HBM is large enough to contain all buffers of the program.

The memory mapping problem for a given program $\mathcal{P}$ with buffers $\mathcal{B}$ is then to decide for each buffer $b\in\mathcal{B}$ whether to allocate space for it in CMEM, and if it is, for which logical time range and at what offset. That is, a solution to the memory mapping problem is a pair of functions $O: \mathcal{B}\rightarrow [0, \mathrm{max\_size}) \cup \{\otimes\}$ and $I: \mathcal{B}\rightarrow [0, T]^2$. The \emph{offset mapping} $O$ assigns each buffer to its offset location in CMEM if it is allocated to it, or it assigns it to a special symbol $\otimes$, denoting that the buffer is to be allocated in HBM. At the same time, the \emph{interval mapping} $I$ assigns each buffer a logical time interval determining the time it is to be allocated in CMEM (and is undefined for buffers not assigned to CMEM). Together, the offset and interval mappings define a memory layout such as the one shown in Figure~\ref{fig:mem_layout}, visualizing the offsets and durations of all the buffers allocated to CMEM.

Another aspect that we model in the memory mapping problem is the data transfer cost to move buffers between HBM and CMEM.
In order for an instruction $\mathcal{I}$ to use a buffer $b$ from CMEM, we need to allocate memory for a time interval that starts long enough before $\mathcal{I}$ to also take transfer time into account (sometimes called \emph{prefetching}). Overall, we want to make sure that time spent on transfer never slows down actual execution time, i.e. that copies are always overlapped fully by computation. To model this, we keep track of the transfer cost of each buffer, which we call its copy \emph{demand value} (typically proportional to its size), as well as the time available at each instruction for copies to be fully overlapped, which we call the \emph{supply value} of a logical step $t$. Now, when allocating a buffer $b$ into CMEM, we need to allocate memory for a logical time interval such that the supply values during the copy duration cover the demand value of $b$.

The construction of $O$ and $I$ needs to adhere to a number of constraints, for instance ensuring that at no point a memory location is oversubscribed to multiple buffers, or that the time range a buffer is allocated to CMEM needs to account for data transfer times. The detailed set of constraints is described in more detail in Appendix~\ref{appendix:game_details}.

\subsection{The \mallocGame\ MDP Environment}\label{subsec:mdp}
We now introduce the memory mapping game MDP, which we refer to as \mallocGame. A direct formulation of the memory mapping problem as MDP could look as follows: We proceed through the buffers in chronological order, and at each step, we define every possible assignment of $O(b)$ and $I(b)$ for the current buffer $b$ as a possible action. The state space captures all possible CMEM states, and state transitions are defined by allocating space for $b$ according to the chosen offset $O(b)$ and interval $I(b)$.

One drawback of this direct formulation is that the action space is extremely large: As every possible value for $O(b)$ and $I(b)$ is a potential action at each step, this accounts for around $10^{12}$ possible decisions at each step. To make the action space more tractable for learning and search, we instead define high-level actions which we call $\Copy$, $\NoCopy$, and $\Drop$ that still capture the key trade-offs the agent needs to make, while allowing for deeper searches and fewer symmetries to learn due the much smaller action space. We now elaborate on our construction of the state space, action space, reward function and dynamics, and then introduce our agent \mallocMuZero\ that plays this game.

\paragraph{State Space}
In the \mallocGame, the player proceeds through each buffer in the order as they chronologically appear in the program, and makes a memory mapping decision. The state at step $t$ is then defined as a tuple $s_t = \langle b_t, O_t, I_t, W_t, \mathcal{B} \rangle$, where $b_t$ is a representation of the current buffer for which a decision needs to be made; $O_t$ is the current offset mapping defined for all previous buffers; $I_t$ is the current interval mapping; $W_t$ is a vector describing the currently available copy supply value at each time step (see Appendix~\ref{appendix:game_details}); and $B$ is the set of all buffers in the program. To represent the state as an input to the agent, We encode the current state of the mapping given by $(O_t, I_t)$ as a two-dimensional binary grid $M_t$, with one axis corresponding to the logical time steps in the program, and the other corresponding to the memory locations in CMEM. A grid cell at coordinate $(t, o)$ is occupied if at time step $t$, the memory location at offset $o$ is occupied, and it is empty, if that memory location is free at $t$. In practice, because of the large size of the grid (up to $32768\times 20000$ in our dataset), the agent typically only sees a down-sampled version of the grid. To identify which buffers are placed at which position in the grid, we also encode $O_t$ as a $t$-length vector with $O_t(b_i)$ at entry $i$. Including the full set of all buffers $\mathcal{B}$ into the state allows the player to plan ahead.

\begin{table}[]
    \centering
    \begin{tabular}{c|l}
        \textbf{Feature} & \textbf{Description} \\
        \hline
        size & Size of the buffer in bytes.\\
        is\_output & Whether the buffer is an output or an operand. \\
        target\_time & Logical time of the instruction using the buffer. \\
        tensor\_id & Id of the corresponding tensor. \\
        alias\_id & Id of the corresponding alias group. \\
        live\_range & Logical time interval for which the buffer is available in the program. \\
        demand & Required data transfer time to move the buffer between HBM and CMEM. \\
        benefit & Estimated speedup if the buffer were placed in CMEM. \\
    \end{tabular}
    \captionsetup{justification=centering}
    \caption{Features used to represent a buffer.}\label{table:buffer_features}
    \label{tab:features}
\end{table}

\begin{figure*}[h!]
    \centering
    \includegraphics[width=0.8\linewidth]{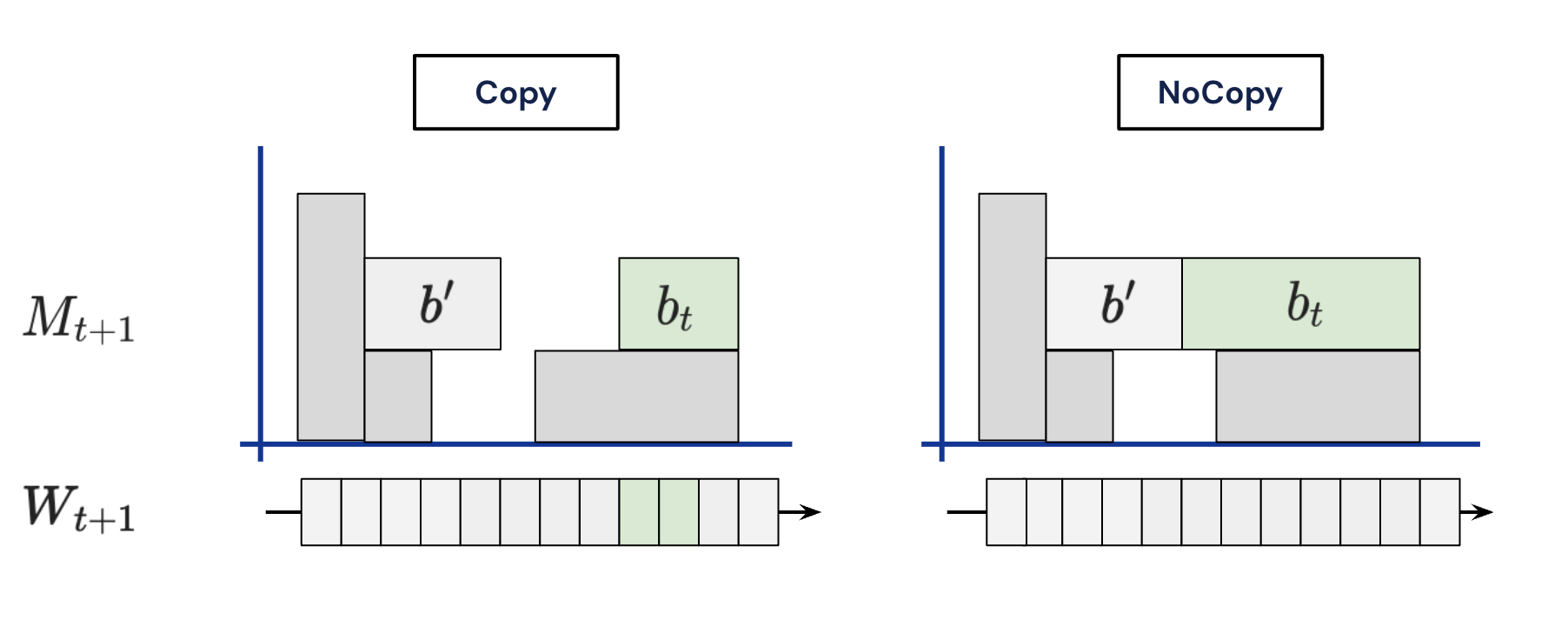}
    \caption{Resource trade-offs represented by actions. The buffer $b_t$ is the current buffer, and $b'$ is a previous buffer already commited to CMEM with $\mathrm{tensor\_id}(b') = \mathrm{tensor\_id}(b_t)$. Choosing $\Copy$ will allocate $b_t$ in CMEM, and reduce available data transfer time in $W_{t+1}$. Choosing $\NoCopy$ will occupy CMEM for a longer time interval, extending the allocation interval of $b'$, but will not impact available data transfer. Choosing $\Drop$ will not impact memory nor data transfer, but may slow down the execution time of instruction $\mathcal{I}_t$.}
    \label{fig:actions}
\end{figure*}

\paragraph{Action Space}
At each step of the game, the player makes a decision for a single buffer $b_t$. The available actions are as follows.
\begin{itemize}
    \item $\Copy$ -- The current buffer $b_t$ is copied from HBM to CMEM. Applying this action will allocate space in CMEM corresponding to the $\mathrm{size}$ of $b_t$, for a time interval such that it covers potential data transfer time between HBM and CMEM.
    \item $\NoCopy$ -- The buffer $b_t$ is placed into CMEM reusing an existing allocation. This action will allocate space in CMEM corresponding to the $\mathrm{size}$ of $b_t$, and extend the interval of the previous allocation of the same tensor, up to the $\mathrm{target\_time}$ of $b_t$. This action is only legal if there is a previous allocation of the tensor of $b_t$, i.e.\ if there is a buffer $b_i$ with $i<t$ and $\mathrm{tensor\_id}(b_i) = \mathrm{tensor\_id}(b_t)$ and $I_t(b_i)$ is an interval that starts before $\mathrm{target\_time}(b_t)$.
    \item $\Drop$ -- The buffer $b_t$ is placed into HBM, and no CMEM allocation is made.
\end{itemize}
These actions capture the trade-offs between the key resources in the memory mapping problem: CMEM space, execution time, and data transfer time. Both $\Copy$ and $\NoCopy$ actions allow the agent to use CMEM space to improve execution times, while $\Drop$ preserves CMEM space for a potential hit on latency. At the same time, $\Copy$ introduces more data transfer between HBM and CMEM, but potentially allocates CMEM space for shorter periods than $\NoCopy$. Figure~\ref{fig:actions} illustrates these trade-offs. For a more formal definition of these actions, and how they translate to offset and interval assignments, we refer to Appendix~\ref{appendix:game_details}.

Note that not all actions are legal in all states. For example, the $\NoCopy$ action can only be applied if there is a matching buffer already in CMEM. One key constraint is the \emph{aliasing constraint}. It imposes that all buffers with the same $\mathrm{alias\_id}$ must either be all assigned to HBM (i.e. applying $\Drop$), or they are all assigned to CMEM (applying either $\Copy$ or $\NoCopy$). More details on the conditions of each action can be found in Appendix Section~\ref{appendix:game_details}. In certain cases, the game can get to a state where none of the actions are legal. In these cases, the game terminates, as the player did not find a feasible memory mapping, and receives a large negative reward. To handle these situations gracefully, we added a backtracking mechanism to our agent, which we introduce in Section~\ref{sec:xlamuzeroagent}.

\paragraph{Reward Function}
In reinforcement learning, finding the right reward function to optimize for is often a challenging task in itself. This also applies to our problem. While the true objective we want to optimize is the latency of the compiled program, measuring it is  unfortunately too costly to include inside the training loop, as compilation can take tens of minutes for each proposed solution. It would also create additional learning challenges, as latency measurement can only be performed after the full memory mapping is decided, and hence it would only provide a single signal at the end of the episodes, which can last over tens of thousands of steps.

To address the limitations of using the latency of the compiled program directly as the reward function, we use a proxy reward: the benefit values for each buffer. These benefit values represent the expected speedup for each buffer if it were placed in fast memory, and are populated themselves by real measurements during a preprocessing step.

The reward function models the incrementally achieved speedup by choosing an action for the current buffer $b_t$. This is a function of the chosen action: For $\Copy$ and $\NoCopy$ actions, the reward is equal to the $\mathrm{benefit}$ of $b_t$. For the $\Drop$ action, the reward is zero, as the HBM speed of the program serves as the baseline. If the player gets into a state where no further actions are legal, the player receives a sufficiently large negative reward such that the total return is less or equal to zero.

One of the key assumptions in this setting is that by maximizing the reward, i.e. the sum of $\mathrm{benefit}$ values of buffers placed into fast memory, we can minimize the latency of the resulting compiled program. This assumption rests on accurate $\mathrm{benefit}$ values that model the real latency speedups as precisely as possible. To obtain accurate $\mathrm{benefit}$ values for each buffer of a given input problem, we make a number of latency measurements using the target hardware (TPUv4(i)). Details of this process is described in Appendix Section~\ref{appendix:game_details}. We indeed observe that our approach performs best when the reward is strongly correlated with latency improvements in the compiled program, and fails as the correlation gets weaker, see Figure~\ref{fig:correlation_plot_joint}.

\paragraph{Dynamics}
Given a state $s_t$, and a chosen action $a_t$, the components of the next state $s_{t+1}=\langle b_{t+1}, O_{t+1}, I_{t+1}, W_{t+1}, \mathcal{B} \rangle$ can be derived as follows:
\begin{itemize}
    \item $b_{t+1}$ -- The next buffer is chosen as the buffer from the set $B$, in the chronological sequence of program instructions.
    \item $O_{t+1}$ -- The offset mapping is updated depending on the action $a_t$: For $\Copy$ and $\NoCopy$ actions, $O_{t+1}(b_{t+1})$ is assigned a valid offset (see Appendix Section~\ref{appendix:game_details} for how this is determined), and for a $\Drop$ action $O_{t+1}(b_{t+1})=\otimes$.
    \item $I_{t+1}$ -- If the action was $\Copy$ or $\NoCopy$, $I_{t+1}(b_t)$ maps to the time interval the buffer occupies memory; and it maps to an empty interval if the action was $\Drop$.
    \item $W_{t+1}$ -- If the action was $\Copy$, we modify $W_t$ by subtracting from it a vector $(u_0, ..., u_T)$ where $u_i$ corresponds to the data transfer time used by the assignment of $b_t$ at time step $i$. How much data transfer time a buffer uses at what time is described in Appendix Section~\ref{appendix:game_details}.
    \item $\mathcal{B}$ -- The set of all buffers $B$ does not change from state to state.
\end{itemize}

\begin{figure*}[ht]
    \centering
    \includegraphics[width=0.8\linewidth]{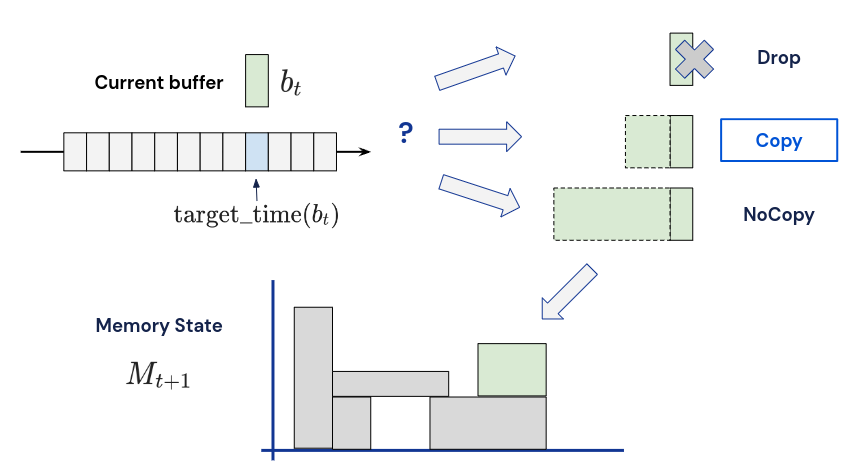}
    \caption{Illustration of one step in the \mallocGame. At state $s_t$, we make a decision about the buffer $b_t$. The $\mathrm{target\_time}$ of $b_t$ points to the logical time of the instruction that uses $b_t$. Each action $a_t \in \{\Copy, \NoCopy, \Drop\}$ defines an offset and a time interval for which CMEM space should be reserved for $b_t$ (in the case of $\Drop$, an empty time interval, and a special offset $\otimes$). The offset, the time interval, and the size of $b_t$ determine which cells in the memory grid $M_{t+1}$ are occupied after applying the action.}
    \label{fig:malloc_game}
\end{figure*}

The game terminates either if the player chooses an invalid action, in which case the game is lost; or the player successfully acts on every buffer in the program, completing the game. The objective is then to find a sequence of valid actions that maximizes the total return of the game. Achieving a high return means that many buffers with high speedup benefits are successfully placed into CMEM, while poor solutions mean slower execution time as important CMEM space is wasted.

\subsection{\mallocMuZero\ Agent}
\label{sec:xlamuzeroagent}

We use deep reinforcement learning to train agents to play the \mallocGame. As training algorithm we use an extension of the MuZero algorithm~\citep{schrittwieser2020mastering}, which we refer to as \mallocMuZero. At a high level, our agent consists of policy, value, reward, and dynamics functions, learned through deep neural networks, and uses a Monte Carlo Tree Search (MCTS) procedure guided by its neural networks to plan ahead in the game. We also introduce the \dropbackup\ mechanism in order to better handle large reward discontinuities as they occur in the \mallocGame.

\subsubsection{Representation encoder}

\begin{figure*}[t]
    \centering
    \includegraphics[width=\linewidth]{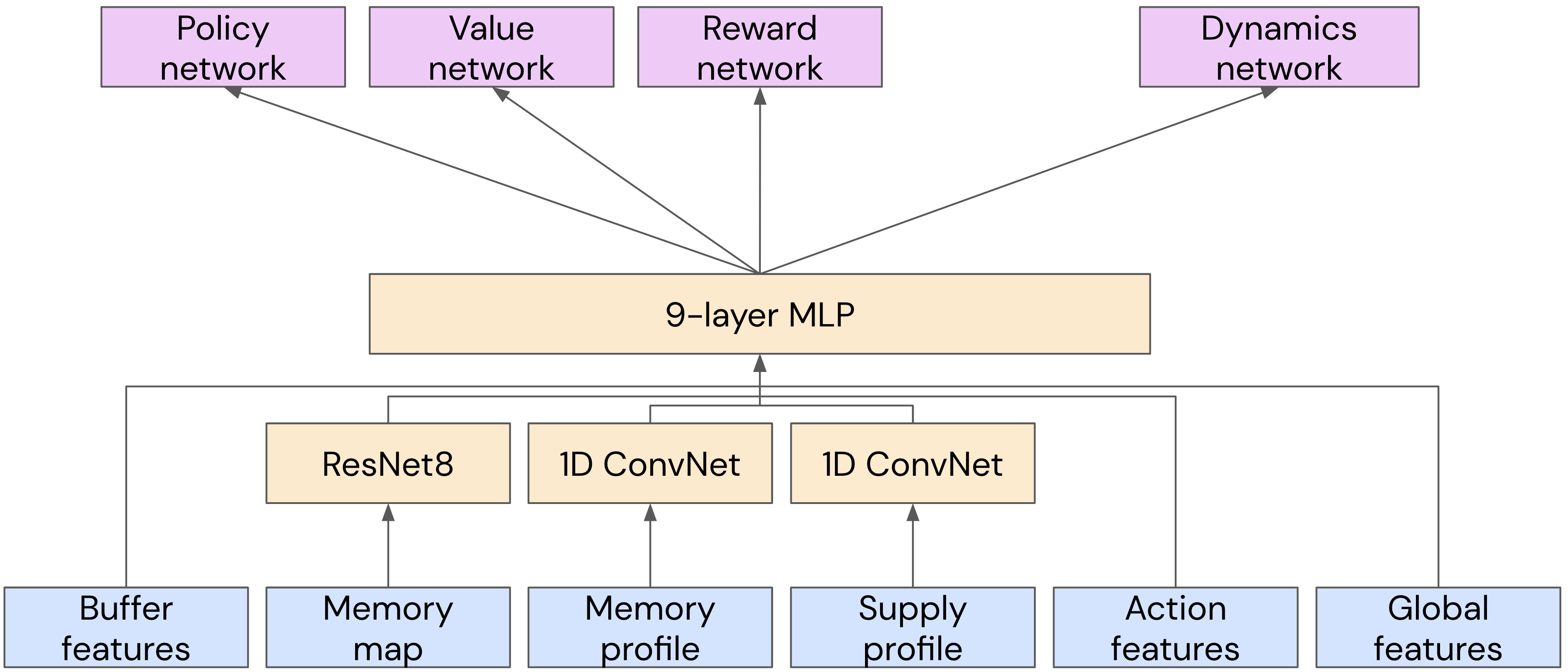}
    \caption{An overview of the \mallocMuZero\ representation network architecture.}
    \label{fig:xla_network_architecture}
\end{figure*}

A key component of the \mallocMuZero\ agent is its representation encoder which translates a representation of the current state $s_t$ to an embedding that is then given to its policy, value, reward, and dynamics networks. The architecture of the representation encoder consists of a combination of ResNet and embedding layers that produce embeddings of raw features, which are then concatenated and passed through a multi-layer perceptron (MLP) to produce the representation embedding, as seen in Figure~\ref{fig:xla_network_architecture}.

To input a state $s_t = \langle b_t, O_t, I_t, W_t, \mathcal{B} \rangle$ to the encoder, we extract relevant features of the state into an initial state representation. This state representation includes:
\begin{itemize}
    \item Buffer features. We include the current buffer $b_t$, as well as the next $k=5$ future buffers $b_{t+1}, ..., b_{t+k}$, as well as the next $l$ buffers $b_i$ that have the same $\mathrm{tensor\_id}$, i.e. $\mathrm{tensor\_id}(b_i) = \mathrm{tensor\_id}(b_t)$. For each buffer, we append the features described in Table~\ref{table:buffer_features}.
    \item Memory map. We provide a fixed size window of the 2-dimensional grid $M_t$ as defined by $(O_t, I_t)$, centered around the $\mathrm{target\_time}$ of $b_t$. Given the large size of this grid in both dimensions, we downsample the grid into a 128x128 binary image.  
    \item Memory profile. For the $\mathrm{target\_time}$ of $b_t$, we also provide a full resolution binary occupancy vector for all memory offsets at the target time.
    \item Supply profile. To provide information about data transfer times, we include a window of the vector $W_t$ centered around the target time of the current job.
    \item Action features. For each of the actions $\Copy$, $\NoCopy$, $\Drop$, we include the legality of the action, the start and end times of its corresponding time interval, and its offset of the corresponding placement.
    \item Global features. We also include global problem features, consisting of the current move number, the current buffer index $t$, and the index of $b_t$ in the order of buffers with the same $\mathrm{alias\_id}$, and the number of buffers remaining with the same $\mathrm{alias\_id}$.
\end{itemize}

These features are concatenated to produce the input to the representation encoder, which then produces a shared embedding used by the policy, value, reward, and dynamics networks of the agent. Each of the policy, value, reward, and dynamics networks consist of MLPs, with different types of outputs: The policy, value, and reward networks produce categorical distributions over actions, values, and rewards respectively, while the dynamics network outputs an embedding for the next state.

\subsubsection{Handling infeasible states}

In the \mallocGame, it is possible to get into \emph{infeasible states}, which do not have any legal actions. This occurs primarily when deciding the placement of a buffer $b$ which must be placed into CMEM due to aliasing constraints (see Section~\ref{subsec:mdp}), but at the same time, can not be placed into CMEM as it would violate memory or bandwidth constraints. Whenever a game reaches such an infeasible state, the game is terminated in a lost state with a return of zero.

The existence of these infeasible states provides a significant challenge to learning. Since the total return of the game resets to zero when getting into an infeasible state, it generally comes with a large negative reward spike that negates the accumulated reward from the episode, meaning large discontinuities in the reward function. Correctly assigning this large negative reward to the offending decisions is very difficult, since decisions leading to the conflict can be arbitrarily far removed from the step where the conflict materializes. Moreover, it can be a combination of actions that collectively lead to a conflict later on, rather than a single incorrect action. Determining which actions lead to infeasibility later on in the game is generally a very hard problem, as it requires reasoning over a combinatorially large number of rollouts and showing that no possible continuation can successfully complete the game.

Hence, to make the task more amenable to learning, we introduce the \dropbackup\ mechanism to our agent. The general idea is based on a key observation about the \mallocGame: If in a state $s_t$, no future buffer shares the same $\mathrm{alias\_id}$ with any already placed job, then dropping (using the $\Drop$ action) all remaining buffers is a valid complete solution to the game. To convince oneself that this holds, we can see that the $\Drop$ action is always valid, unless it violates the alias group constraint. By imposing that there is no intersecting alias groups between past and future, we can then assign all future uses the $\Drop$ action without worrying about infeasibility.

We utilize this observation as follows. Instead of playing a single game, our agent maintains two game trajectories at the same time: The current main game trajectory, as well as a \emph{backup} trajectory that contains a prefix of the main trajectory that can be extended to a full solution. Whenever the agent moves into an infeasible state by choosing some action $a_t$ at some state $s_t$, we reset the game to the backup trajectory, apply $\Drop$ actions to all jobs with the same $\mathrm{alias\_id}$ as $b_t$, and save a new backup. In this way, the agent can keep its past progress even when encountering an infeasible state in the middle of the game, and avoids large negative rewards that reset the return to zero. Resetting to the backup state still provides a negative reward signal, though it is more local, making credit assignment much more tractable.

\section{Experiments}
\label{sec:results}

We now present our experimental results for solving the memory mapping problem using our agent \mallocMuZero\, and compare it to the current solver present in the XLA library~\citep{xla}. We describe the experimental setup, and present our main results on the achieved latency on a number of realistic machine learning workloads, including the XLA TPU benchmark suite used for benchmarking compiler changes in XLA, as well as a number of workloads that have a large resource footprint at Alphabet. Finally, we also provide a set of investigative studies to shed more light on the performance of the \mallocMuZero\ agent.

\subsection{Experimental Setup}
\label{subsec:experimental_setup}

\paragraph{Production \mallocMuZero}
Our approach works \emph{offline} from the XLA compilation process, i.e. our agent is not trained and run during the compilation process, but instead is designed to be run separately in parallel or in advance. A common use case for this offline setup is to target high-importance workloads for which higher resource efficiency or smaller latency are extremely desirable, and taking this additional step during compilation is worth it. Another advantage of this offline approach is that we do not need to fully replace the default heuristics inside the XLA compiler that have been tuned over years of engineering experience. Instead, we can make use of the best of both worlds, and run both our \mallocMuZero\ agent and the XLA compiler heuristics in parallel, and take the best result from both. This would be the preferred setup used in production, and we call this version of our approach the \prodmallocMuZero\ agent. In practice, this is a realistic setting in which our approach could be integrated with XLA, keeping the reliability of a well-understood heuristic, while still reaping the benefits of better solutions found by our agent.

\paragraph{Dataset}
We evaluate our approach on a set of 52 machine learning workloads that are part of the XLA benchmark, as well as 8 additional workloads with a high resource footprint sampled from across Alphabet, including a version of the AlphaTensor~\citep{alphatensor} model. The benchmark workloads are used in the development of the XLA compiler to track improvements and regressions, and cover a broad range of architectures and applications. This set also covers a large range in terms of problem size, ranging from 169 buffers to assign in the smallest instance, up to 16490 buffers in the largest. Appendix~\ref{appendix:dataset} includes more details and statistics about the dataset.

\paragraph{Baselines}
As a baseline, we use the XLA compiler invoked with default parameters. Unlike our approach here, the memory mapping solver in the XLA compiler is not based on solving an explicitly defined optimization problem, but rather uses a set of heuristics designed and refined over years by domain experts. As the XLA compiler is still constantly evolving, we use a recent version of it, including changes up until July 2022. We compare both the stand-alone \mallocMuZero\ as well as the hybrid version \prodmallocMuZero\ to the XLA compiler in terms of latency of the compiled programs.

In addition to comparing how we perform compared to the XLA compiler on producing efficient memory mappings, we also want to understand the performance of \mallocMuZero\ as an optimizer to the \mallocGame. As the XLA compiler does not optimize directly for the \mallocGame\ formulation, we instead compare it against another black-box optimization approach. For this purpose, we implemented an evolutionary search approach based on~\citep{salimans2017evolution}, which performs a guided search across the search space of actions to play in the \mallocGame.

\paragraph{Metrics} 
The primary metric we want to optimize is the execution latency of the compiled program. To measure this, we modified the XLA compiler with an optional flag to use memory mapping solutions generated by our agent during compilation. For a given program, we then compile it using XLA for TPUv4i as hardware target, and run the compiled program on a machine with a single TPUv4i chip. The latency is measured as the total end-to-end time of the program spent on the TPU device, i.e.\ we exclude any time spent on the host machine (CPU). This provides a more accurate measurement of the effect of the memory mapping, since it only affects the TPU, and allows us to avoid typically noisy CPU latency measurements. 

To compare against the baseline, we use the relative speedup measure defined as follows:
\begin{equation*}
    \mathrm{speedup}=\frac{\mathrm{latency_{baseline}}}{\mathrm{latency_{\mallocMuZero}}}
\end{equation*}
Speedup values above 1 correspond to our agent finding faster solutions than the baseline, while values below 1 mean that the baseline solution is faster. We sometimes report speedup values as percentage improvements, i.e. a $x\%$ speedup corresponds to a speedup value of $1+\frac{x}{100}$. Note that while the stand-alone \mallocMuZero\ agent does not always reach speedup values of $1$, the benefit of \prodmallocMuZero\ is that it is guaranteed to score $1$ or higher on the speedup metric.

\paragraph{Training and Evaluation Setup}
For each of the 60 workloads in the dataset, we extract the corresponding memory mapping problem from the XLA compiler into a \mallocGame. We train a \mallocMuZero\ agent from scratch for each game, for a training period of up to 24 hours. Training hyperparameters are given in Appendix~\ref{appendix:hypers}. We take the best solution found by each agent with respect to achieved rewards, and evaluate it by supplying the memory mapping solution back to the XLA compiler. We then measure its execution latency of the compiled program on a single TPUv4i chip. We do the same for the baseline, only using the compiled program from the default XLA compilation process.

\subsection{Results}
\begin{figure*}[t!]
\centering
    \begin{tabular}{cc}
         \includegraphics[width=0.45\linewidth]{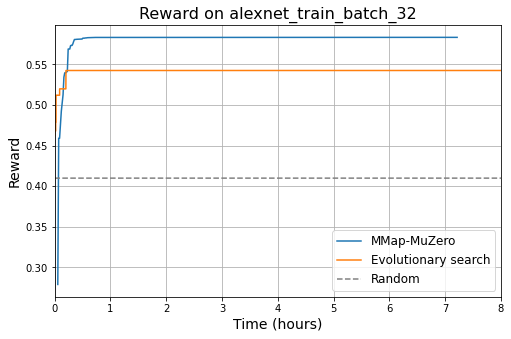} & \includegraphics[width=0.45\linewidth]{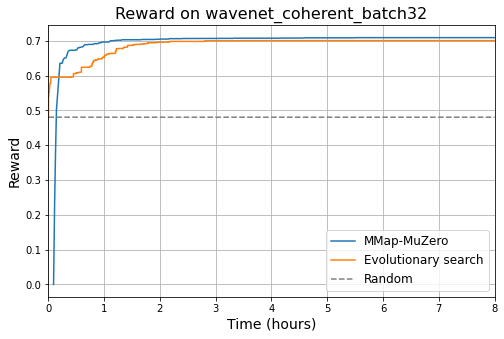} \\
         \includegraphics[width=0.45\linewidth]{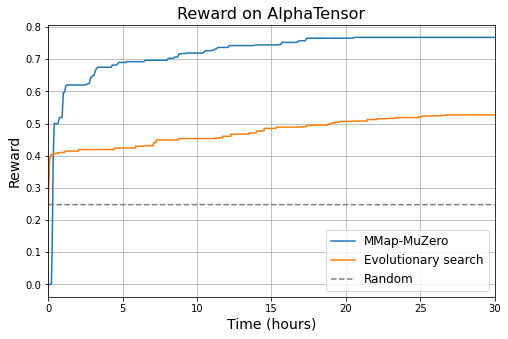} & \includegraphics[width=0.45\linewidth]{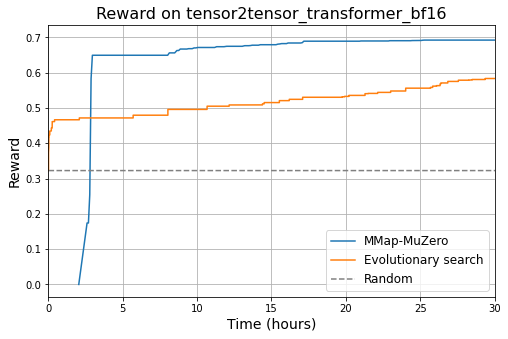}
    \end{tabular}
    \caption{Reward achieved by \mallocMuZero, evolutionary search, and random policy across time on the same hardware budget.}\label{fig:reward_curves}
\end{figure*}

\paragraph{Rewards}
We start with investigating the performance of \mallocMuZero\ as an optimizer of the \mallocGame, compared to an evolutionary search baseline (ES) as well as a random baseline that takes legal actions at random. We measure the rewards achieved by both approaches on a subset of \mallocGame\ instances from our dataset containing problems of different sizes. Given the different computational cost of each optimization step between evolutionary search and training our reinforcement learning agent, we compare performance achieved against a fixed time budget on the same hardware. Figure~\ref{fig:reward_curves} shows the reward curves on four problem instances, and Table~\ref{table:rewards}. Overall, \mallocMuZero\ quickly achieves higher rewards than the evolutionary search approach after a short time period on all problems. Especially when the problem size grows large, we see reinforcement learning scaling better than evolutionary search, as the search space grows too large to explore efficiently through local mutations.

\begin{table}[h]
\centering
\begin{tabular}{c|c|c|c|c}
 \textbf{Model} & \textbf{Size} & \textbf{\mallocMuZero} & \textbf{ES} & \textbf{Random} \\
 \hline
 alexnet\_train\_batch\_32 & 300 & \textbf{0.5834} & 0.5426 & 0.4100 \\
 wavenet\_coherent\_batch32 & 3020 & \textbf{0.7099} & 0.7004 & 0.4804 \\
 AlphaTensor & 9084 & \textbf{0.7680} & 0.5357 & 0.2481 \\
 tensor2tensor\_transformer\_bf16 & 9888 & \textbf{0.7002} & 0.6263 & 0.3227 \\
\end{tabular}
\captionsetup{justification=centering}
\caption{Final reward achieved by \mallocMuZero, evolutionary search (ES), and choosing actions randomly.}\label{table:rewards}
\end{table}

\paragraph{Latency}
We now present results on the latency speedups found by our approach. Across the full dataset, \mallocMuZero\ improves end-to-end latency for 33 out of 60 programs, with a speedup of up to $87\%$ on the workload with the best relative performance. As a stand-alone agent, \mallocMuZero\ achieves an average speedup of $0.59\%$, while our hybrid approach \prodmallocMuZero\ achieves an average speedup of $4.05\%$ across all workloads. Table~\ref{table:speedup} shows the aggregate speedups achieved by both agents, and Table~\ref{table:top_speedup} shows the instances with the best and worst relative performance of \mallocMuZero. The full breakdown of the speedup found for each individual model can be found in Appendix~\ref{appendix:dataset}.

The latency results for \mallocMuZero\ also show that there are workloads for which our approach underperforms the heuristics in XLA, and where we do not find any speedup compared to the default XLA compiler. While we do not expect all workloads to be able to be sped up, as not all workloads are bottlenecked by memory mapping, problems where our agent performs significantly worse than the baseline deserve a closer analysis. We provide some investigative studies on this in Section~\ref{subsec:analysis}.

\begin{table}[h!]
    \centering
    \begin{tabular}{c|c|c|c|c}
        \textbf{Agent} & \textbf{Mean} & \textbf{Max} & \textbf{Min} & \textbf{\# of models improved}\\
        \hline
        \mallocMuZero & 1.0059 & 1.8787 & 0.4853 & 33/60 \\
        \prodmallocMuZero & \textbf{1.0405} & \textbf{1.8787} & \textbf{1.0} & \textbf{33/60}
    \end{tabular}
    \captionsetup{justification=centering}
    \caption{Mean, minimum, and maximum speedup values by our agents.}\label{table:speedup}
\end{table}

\begin{table}[h]
    \centering
    \begin{tabular}{c|c|c|c}
        \textbf{Models with largest speedup} & \textbf{Speedup} & \textbf{Models with lowest speedup} & \textbf{Speedup}\\
        \hline
        inference\_lstm & 1.8787 & rnn\_7 & 0.4853 \\
        wavenet\_coherent\_batch32 & 1.3335 & rnn\_6 & 0.5482 \\
        alexnet\_train\_batch\_32 & 1.1709 & mlperf\_nmt\_1\_shard\_batch\_64 & 0.7892 \\
        mnasnet\_b1\_batch\_128 & 1.0960 & rnn\_3 & 0.8541 \\
        inference\_resnet & 1.0893 & tensor2tensor\_transformer & 0.9125 \\
        AlphaTensor & 1.0578 & rnn\_2 & 0.9343
    \end{tabular}
    \captionsetup{justification=centering}
    \caption{Models with most and least speedup from \mallocMuZero.}\label{table:top_speedup}
\end{table}

\begin{figure*}[t!]
    \centering
    \includegraphics[width=1.0\linewidth]{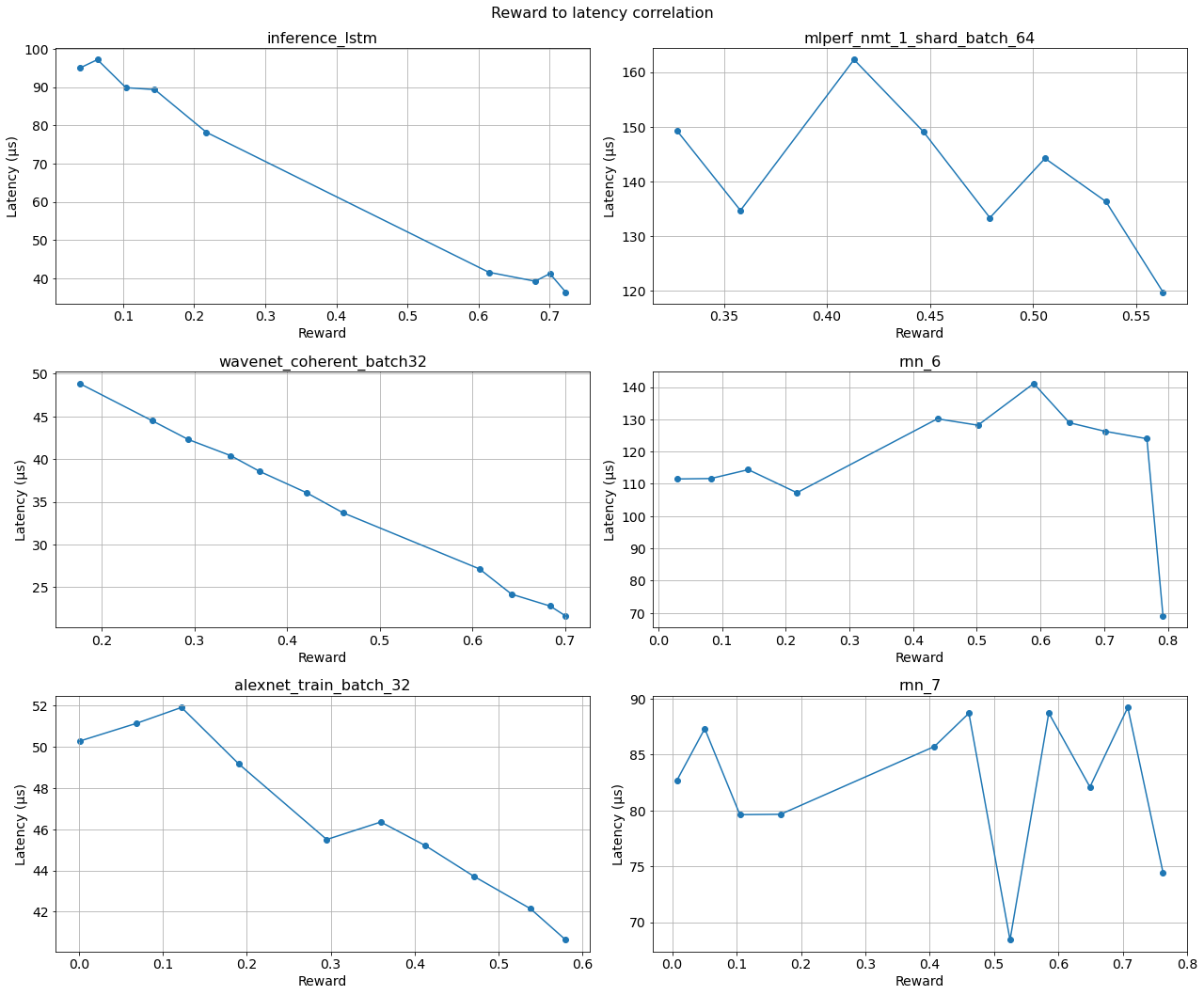}
    \caption{Reward to latency correlation plots for the instances with top 3 (on the left) and bottom 3 (on the right) speedups. Having a strong inverse correlation allows \mallocMuZero\ to optimize for the reward which translates into a latency speedup.}
    \label{fig:correlation_plot_joint}
\end{figure*}

\begin{table}[h]
    \centering
    \begin{tabular}{c|c|c}
        \textbf{Model} & \textbf{Speedup} & \textbf{Reward correlation}\\
        \hline
        inference\_lstm & 1.8787 & -0.9957 \\
        wavenet\_coherent\_batch32 & 1.3335 & -0.9994 \\
        alexnet\_train\_batch\_32 & 1.1709 & -0.9581 \\
        \hline
        rnn\_7 & 0.4853 & -0.0774 \\
        rnn\_6 & 0.5482 & 0.0299 \\
        mlperf\_nmt\_1\_shard\_batch\_64 & 0.7892 & -0.5422
    \end{tabular}
    \captionsetup{justification=centering}
    \caption{Correlation between reward and latency for workloads with best and worst \mallocMuZero\ performance.} \label{table:correlation}
\end{table}

\subsection{Analysis}\label{subsec:analysis}

\paragraph{Correlation between reward and latency}
As discussed in Section~\ref{subsec:mdp}, our reward function is only a proxy for the real latency of the compiled program. This discrepancy can be one of the explanations for the poor performance of \mallocMuZero\  on certain instances. To understand if this is indeed the case, we studied the correlation between the true objective (latency) and our reward function for both the top-3 best-performing instances (largest speedups) and the bottom-3 instances (smallest speedups). For each instance, we sample ten different solutions from different stages of a \mallocMuZero\ training run, with different reward values achieved. We then measure the latency of the corresponding compiled program for each of the solutions, and calculate the Pearson correlation coefficient between reward achieved and latency measured.

In Figure~\ref{fig:correlation_plot_joint} and Table~\ref{table:correlation}, for the problems where \mallocMuZero\ performs well, we can see a strong negative correlation between rewards and latency, which is ideal for the reward function: Higher rewards should lead to lower latency. On the other hand, on the instances where \mallocMuZero\ fails to find fast memory mappings, this relationship does not hold: Reward and latency are not well correlated, and optimising for reward does not mean that latency is minimised. This emphasises how crucial finding a good reward function is, and that our choice of reward function is not working well in all cases. More optimistically, this also suggests that our results can be significantly improved by improving our reward function, for example, by using a more realistic benefit calculation or more accurate cost models for latency.

\paragraph{Ablation study}
We conducted an ablation study to understand the contribution of the MCTS and learning components of our \mallocMuZero\ agent. To do this, we performed two training runs: (1) A run without learning, which only performs pure MCTS using the true dynamics of the \mallocGame\ instead of the learned dynamics model (cf.~\citep{muzero}); and (2) a run with learning, but with MCTS disabled. We compare both runs with the full \mallocMuZero\ agent.

\begin{figure*}[h]
    \centering
    \includegraphics[width=0.8\linewidth]{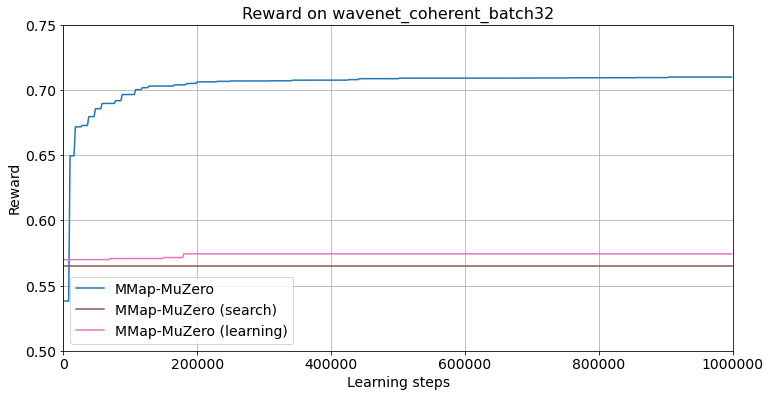}
    \caption{Performance of \mallocMuZero\ when it only uses its search or learning components.}
    \label{fig:ablation_plot}
\end{figure*}

As we can see from Figure~\ref{fig:ablation_plot}, the best performance curve is achieved in the full setting, with both search and learning components. This highlights the importance of both search and learning to achieving the best results.

\paragraph{Memory layout comparison}

\begin{figure*}[h!]
    \centering
    \begin{subfigure}[b]{0.49\linewidth}
        \centering
        \includegraphics[width=\linewidth]{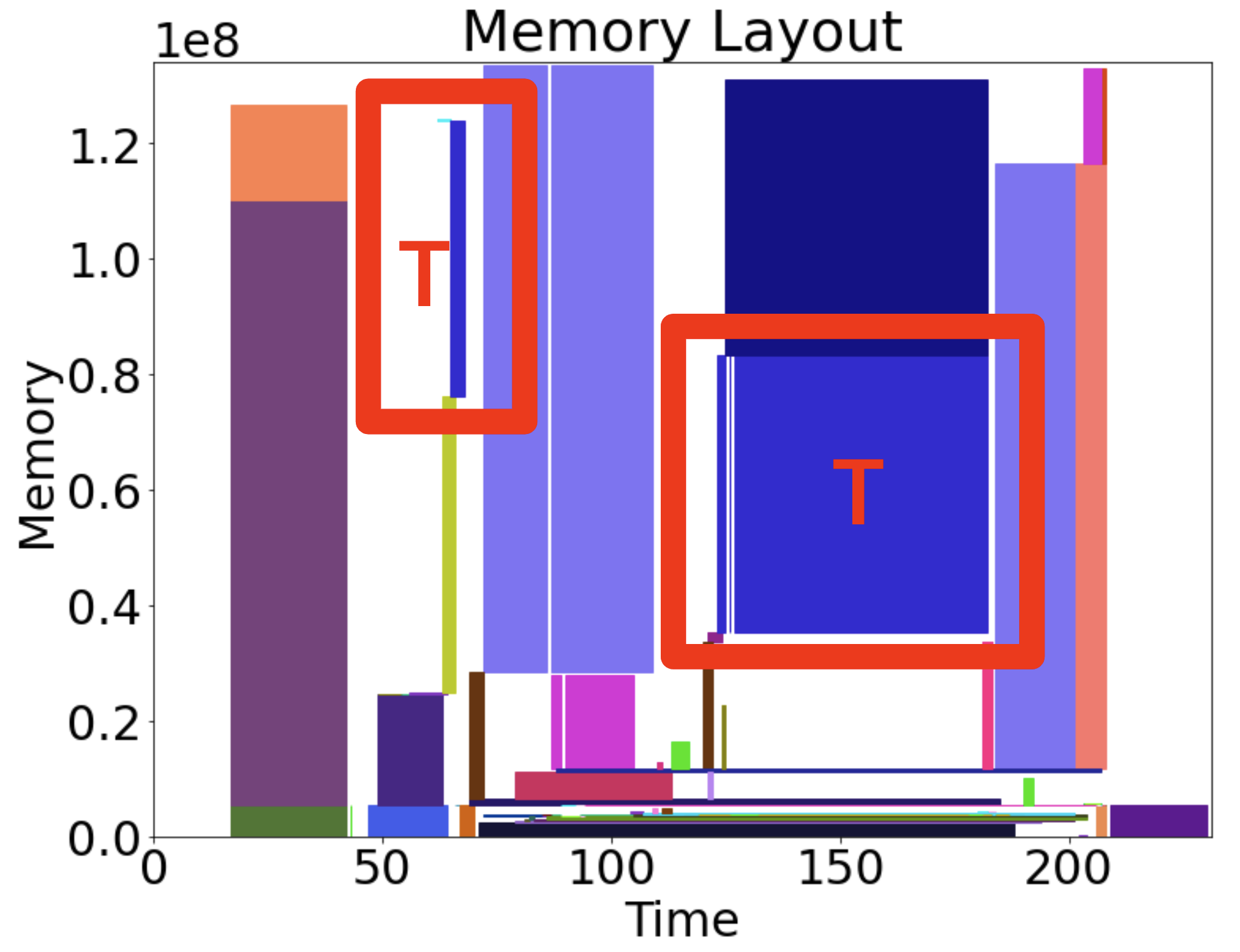}
        \caption{\mallocMuZero}
    \end{subfigure}
    \begin{subfigure}[b]{0.49\linewidth}
        \centering
        \includegraphics[width=\linewidth]{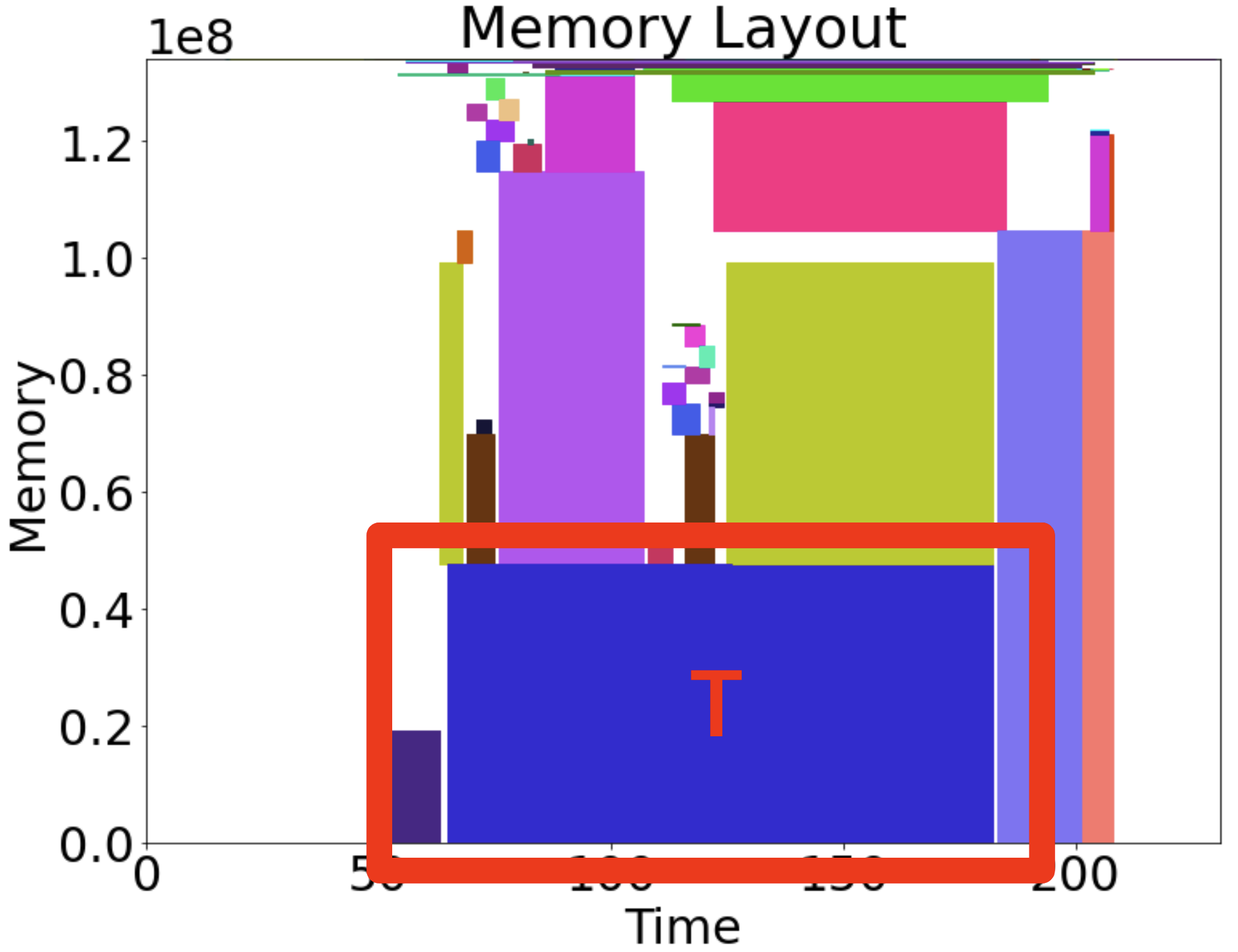}
        \caption{XLA production heuristic}
    \end{subfigure}
    \caption{Memory layouts for \emph{alexnet\_train\_batch\_32} by \mallocMuZero\ (left) and the XLA production heuristic (right). Each rectangle represents an assignment of a buffer into fast memory, buffers with the same colour correspond to the same tensor. Highlighted is a tensor $T$ which \mallocMuZero\ loads into CMEM multiple times, leaving space for other buffers in between uses, while the production heuristic keeps $T$ in CMEM for its whole lifetime. The layout found by $\mallocMuZero$ achieves a latency improvement of 17\%.}
    \label{fig:memory_layout}
\end{figure*}

We also inspected the memory layouts produced by \mallocMuZero\ to understand qualitative differences to the XLA heuristic. One illustrative example is shown in Figure~\ref{fig:memory_layout}, which shows the memory layouts produced for the \emph{alexnet\_train\_batch\_32} model, for which \mallocMuZero\ finds a $17\%$ speedup compared to the XLA heuristic. We can see the two approaches making drastically different allocation decisions, highlighting that our agent discovers highly performant memory mappings from scratch, without expert guidance. One interesting difference here is how the agent makes more frequent use of offloading tensors out of fast memory if they are are not needed for a long period, thereby freeing up space in between uses. As an exmaple, this can be seen for the highlighted tensor marked $T$: In the memory layout found by \mallocMuZero, $T$ is loaded into and and evicted from CMEM multiple times, leaving space for other tensors. In the production heuristic, $T$ is loaded once into CMEM, and never leaves it within its lifetime, and not enough space is left for other important tensors. This highlights again the complexity of this problem, and how a learned agent can find solution approaches that either have not been considered, or are too complex to formulate as a general heuristic.

\section{Discussion}
\label{sec:discussion}
In this paper, we presented a deep reinforcement learning approach to solve the memory mapping problem occurring in the XLA compilation process. Solving the memory mapping step well is crucial to produce fast, low-latency compiled programs, as memory access is often a key bottleneck. We defined this problem as a Markov decision process in the form of a single-player game, the \mallocGame. We introduced our agent, \mallocMuZero, an extension of the MuZero \citep{schrittwieser2020mastering} agent that plays this game and comprised a novel representation network and the \dropbackup\ mechanism to avoid infeasible states.

On a set of realistic ML workloads, including $52$ from the XLA benchmark, and $8$ high-impact workloads from the Alphabet's fleet, we improved the execution times of $33$ programs. Our agent \mallocMuZero\ achieved an overall average speedup of $0.59\%$ with a maximum speedup of $87\%$ against the default XLA compiler. On the AlphaTensor \citep{alphatensor} model, we sped up execution by $5.78\%$. We also introduced a hybrid agent, \prodmallocMuZero, which combines \mallocMuZero\ with the production baseline, yielding an agent suitable for productionization. The hybrid agent improved upon both the XLA baseline as well as \mallocMuZero\ to yield an average speedup of $4.05\%$, indicating the large potential of this approach.

We also ran a set of investigative studies to better understand the performance of the agent and found that the correlation between the reward function and execution time is critical to yielding improved performance with our agent. This adds validation to our case that deep reinforcement learning approaches are a powerful tool to model and solve complex combinatorial problems.

\bibliographystyle{abbrvnat}
\nobibliography*
\bibliography{refs}


%







\clearpage
\appendix
\onecolumn

\section{Details of \mallocGame}
\label{appendix:game_details}
In this section, we provide further technical details of the \mallocGame.

\paragraph{Aliasing}
XLA supports aliasing\footnote{https://www.tensorflow.org/xla/aliasing}, which allows multiple expressions to refer to the same underlying memory location. In the \mallocGame, we model these specifications by allowing multiple buffers to be placed in the same \emph{alias group}. Each alias group is identified by an id, and the memory allocations made for buffers in the same alias group must be at the same offset:
\begin{equation}\label{eq:aliasing}
    \forall b_1, b_2: \mathrm{alias\_id}(b_1) = \mathrm{alias\_id}(b_2) \Rightarrow O(b_1) = O(b_2).
\end{equation}

\paragraph{Data transfer between fast and slow memory}
One key resource that needs to be managed when playing the \mallocGame\ is the data transfer time between memories, e.g. between HBM and CMEM on TPUv4(i). Given a finite transfer bandwidth, copying a buffer from HBM to CMEM takes time proportional to the size of the buffer that is transferred. Hence, in order for an instruction $\mathcal{I}$ to use a buffer $b$ from CMEM, we need to allocate CMEM memory for a time interval that starts long enough before $\mathcal{I}$ to also take transfer time into account (sometimes called \emph{prefetching}). Overall, we want to make sure that time spent on transfer never slows down actual execution time, i.e. that copies are always overlapped fully by computation.

In the \mallocGame, we model this as follows. Each buffer $b$ has a specified copy \emph{demand value}, which describes the amount of time it takes to copy $b$ between HBM and CMEM at the fixed maximum transfer bandwidth of the hardware. In our case, we use the size of the buffer multiplied by a hardware-specific bandwidth constant as the demand value for each buffer. Furthermore, for each logical time step of the program, i.e. each instruction, we assign it a \emph{supply value}, describing the amount of time the execution of the program spends on that instruction. Now, when allocating a buffer $b$ into CMEM, we need to allocate memory for a (logical) time interval such that the supply values during the copy duration cover the demand value of $b$. Formally, let us define the \emph{copy interval} of a buffer $b$ and its allocation interval $I(b) = [s, e]$ as follows:
\begin{itemize}
    \item $\mathrm{copy}(b) := [s, \mathrm{target\_time}(b))$ if $b$ is an \emph{input} buffer placed using a \Copy\ action.
    \item $\mathrm{copy}(b) := (\mathrm{target\_time}(b), e]$ if $b$ is an \emph{output} buffer placed using a \Copy\ action.
    \item $\mathrm{copy}(b)$ is the empty interval for any buffer placed with \NoCopy\ or \Drop\ actions.
\end{itemize}
We then require that
    \begin{equation}\label{eq:supplydemand}
        \sum_{t\in \mathrm{copy}(b)} \mathrm{supply}(\mathcal{I}_t) \geq \mathrm{demand}(b).
    \end{equation}

Note that the supply values of instructions are updated throughout the \mallocGame, described by the dynamics of the game. In addition to the above constraints, we also impose that there is only a single buffer being copied between memories at any given point in time. This is modeled in the \mallocGame\ by imposing that the copy intervals of any two buffers have no internal intersections.
    \begin{equation}\label{eq:time_nonintersect}
        \forall b_1, b_2: |\mathrm{copy}(b_1) \cap \mathrm{copy}(b_2)| \leq 1. 
    \end{equation}
This restriction aims to make sure that any copy of a buffer proceeds with the undivided maximum bandwidth available on the hardware. While this does not always turn out to be the case in practice on the real hardware, it is a workable approximation to ensure that copy time does not slow down the critical path of execution.

\paragraph{Assigning allocation intervals and offsets}
Choosing the \Copy\ or \NoCopy\ action for a given buffer $b$ means to allocate it in fast memory for some time interval $I(b)$ at some offset $O(b)$. We now describe how $I(b)$ and $O(b)$ are determined in the \mallocGame.

For \Copy:
\begin{itemize}
    \item If $b$ is an \emph{input}, then $I(b) = [s, \mathrm{target\_time}(b)]$ where $s$ is the latest logical time step, such that Equations~\ref{eq:supplydemand} and \ref{eq:time_nonintersect} are satisfied.
    \item If $b$ is an \emph{output}, then $I(b) = [\mathrm{target\_time}(b), e]$ where $e$ is the earliest logical time step, such that Equations~\ref{eq:supplydemand} and \ref{eq:time_nonintersect} are satisfied.
    \item $O(b)$ is chosen as the lowest offset, such that the CMEM offset range $[O(b), O(b) + \mathrm{size}(b))$ is fully available across the full time interval $I(b)$, and Equation~\ref{eq:aliasing} is satisfied.
\end{itemize}

For \NoCopy:
\begin{itemize}
    \item If $b$ is an \emph{input}, then $I(b) = (s, \mathrm{target\_time}(b)]$ where $s$ is the latest logical time step that lies within a time interval assigned to a buffer $b'$ with $\mathrm{tensor\_id}(b) = \mathrm{tensor\_id}(b')$.
    \item If $b$ is an \emph{output}, then $I(b) = \mathrm{live\_range}(b)$.
    \item $O(b)$ is chosen in the same way as for \Copy.
\end{itemize}

\paragraph{Benefit and supply values}
To model changes to the execution time due to actions taken in the \mallocGame, and to model the time taken by data transfer faithfully, we depend on accurate values for populating the benefit values and supply values. Two common approaches to determine execution times in optimization problems are (1) to use a mathematical \emph{cost model} that approximates the latency of operations using features (such as the size of the inputs, the type of instruction, etc.); or (2) to use real hardware measurements. Both approaches generally have distinct pros and cons, with a cost model typically being cheap to evaluate, but less accurate, and hardware measurements being expensive, but more accurate. In our work, we tried both approaches initially, but settled on a simplified hardware measurement approach that yields generally good enough approximations without being prohibitively expensive.

In our approach, we measure the execution time of each instruction of the program individually under a diverse set of memory assignments. For each instruction $\mathcal{I}$ with inputs $i_1, ..., i_n$ and outputs $o_1, ..., o_m$ we measure the execution time of $I$ for every combination of assigning any subset of $i_1, ..., i_n$ and $o_1, ..., o_m$ to CMEM. To limit the total number of measurements, we choose to vary only the largest $8$ inputs/outputs if $n+m>8$. This results in $\min(2^8, 2^{n+m})$ many measurements per instruction in the program. Let $L_\mathcal{I}(\{b_1, ..., b_k\})$ denote the measured execution time of instruction $I$ when buffers $\{b_1, ..., b_k\}$ were allocated to CMEM.

Given these latency measurements for each instruction, we calculate and update benefits and supply values as follows:
\begin{itemize}
    \item The initial benefit value of each buffer $b\in B(\mathcal{I})$ is set to $L_\mathcal{I}(\{\}) - L_\mathcal{I}(\{b\})$; i.e. the latency delta between the full HBM allocation of $B(\mathcal{I})$, and just putting $b$ into CMEM.
    \item The initial supply value of $\mathcal{I}$ is set to $L_\mathcal{I}(B(\mathcal{I}))$, where $B(\mathcal{I})$ denotes the full set of input and outputs of $\mathcal{I}$. This is generally an underestimate of the actual execution time of $\mathcal{I}$ to make sure data transfer does not impact the critical path.
    \item At each step, when considering a buffer $b$ of an instruction $\mathcal{I}$, the benefit of $b$ is updated to $L_\mathcal{I}(B') - L_\mathcal{I}(B'+\{b\})$, where $B'$ denotes the set of buffers of $\mathcal{I}$ currently already allocated to CMEM.
\end{itemize}

\section{Dataset and full results}\label{appendix:dataset}
\begin{figure*}[h!]
    \centering
    \includegraphics[width=0.8\linewidth]{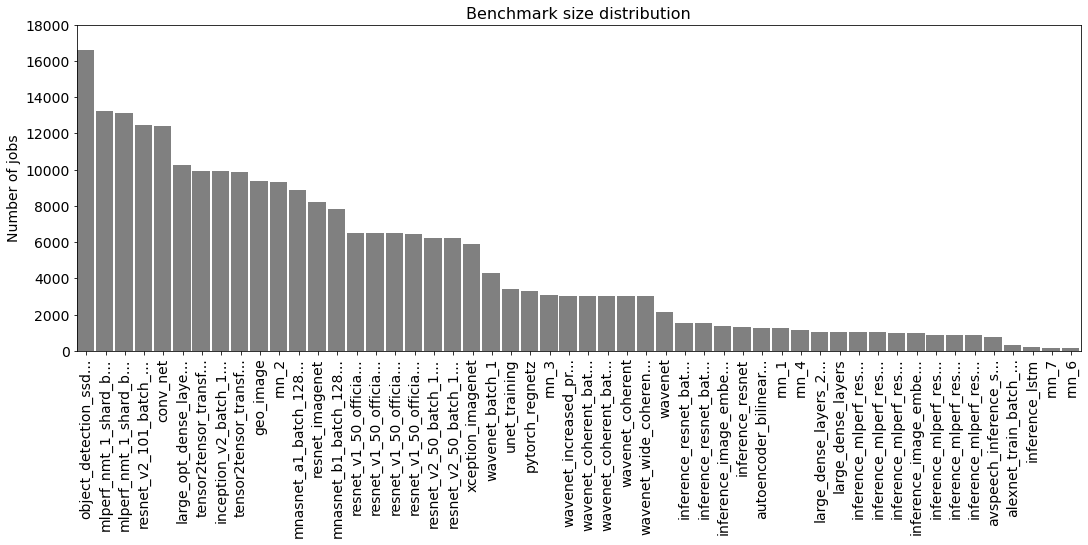}
    \caption{The number of buffers for each workload in the benchmark set.}
    \label{fig:num_jobs_plot}
\end{figure*}

\begin{figure*}[h!]
    \centering
    \includegraphics[width=0.8\linewidth]{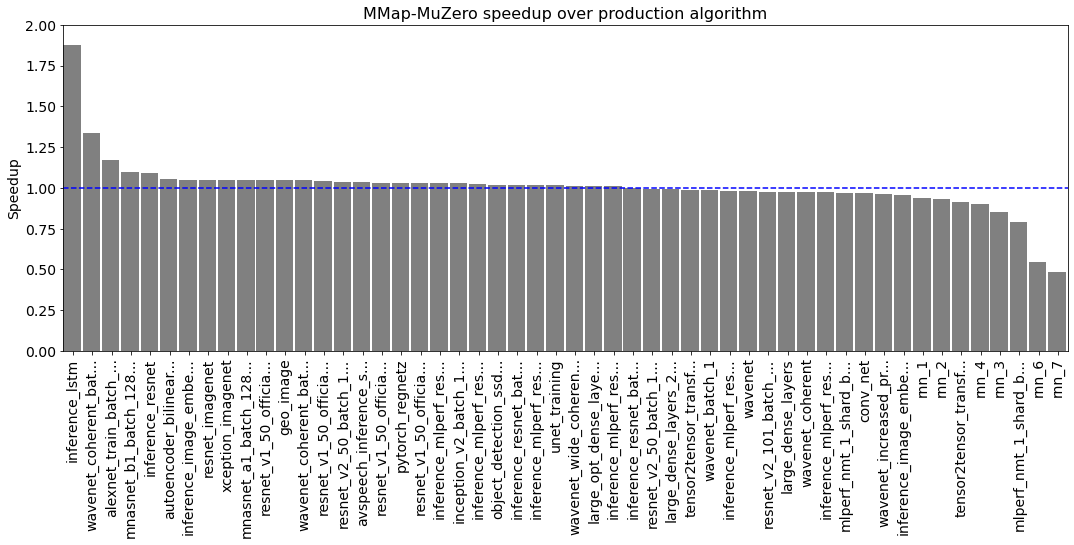}
    \caption{Speedup values for each workload.}
    \label{fig:speedup_plot}
\end{figure*}

We list all XLA workloads we used for evaluation of \mallocMuZero, along with their problem size (Figure~\ref{fig:num_jobs_plot}), and the achieved speedup (Figure~\ref{fig:speedup_plot}).

\section{Hyperparameters}\label{appendix:hypers}

We use the following fixed hyperparameters for training \mallocMuZero\ on all models:

\begin{table}[h]
    \centering
    \begin{tabular}{c|c|c}
        \textbf{Hyperparameter} & \textbf{Value} & \textbf{Description} \\
        \hline
        discount\_factor & 0.9999 & Discount factor for episode rewards \\
        num\_mcts\_simulations & 400 & Number of MCTS simulations before action selection \\
        init\_temperature & 1.0 & Initial temperature for the action selection \\
        temperature\_decay\_steps & 400000 & Number of steps when the temperature is decayed \\
        final\_temperature & 0.2 & Final temperature after decay \\
        noise\_fraction & 0.25 & Fraction of Dirichlet noise to mix in with prior in MCTS \\
        noise\_alpha & 0.03 & Dirichlet noise parameter \\
        num\_training\_steps & 1000000 & Number of optimization steps \\
        optimizer & adam & Optimizer \\
        batch\_size & 512 & Optimization batch size \\
        lr & 0.0002 & Learning rate \\
        weight\_decay & 0.0001 & Weight decay applied to parameters \\
        replay\_size & 20000 & Replay buffer size \\
        reanalyse\_fraction & 0.5 & Fraction of training data from Reanalyse \\ 
    \end{tabular}
    \captionsetup{justification=centering}
    \caption{\mallocMuZero\ hyperparameters.} \label{table:hyperparameters}
\end{table}

\end{document}